\documentclass[superscriptaddress,nofootinbib,twocolumn,aps]{revtex4-2}
\usepackage{ulem}
\usepackage{epsfig,amsmath,amssymb,amsfonts,color,algorithm,algcompatible,amsthm,bm,braket}
\usepackage{multirow}
\usepackage[table,xcdraw]{xcolor}
\usepackage[colorinlistoftodos,prependcaption]{todonotes}
\usepackage{threeparttable}
\usepackage{subfigure}

\newcommand{\sgn}[1]{\text{sgn}\left\lbrack#1\right\rbrack}
\newcommand{\data}[0]{|{\rm Data}\rangle}
\newcommand{\fig}[4]{
\begin{figure*}[ht]
\centering
\includegraphics[#4]{#1}
\caption{#2}
\label{#3}
\end{figure*}
}
\newcommand{\fighalf}[4]{
\begin{figure}
\centering
\includegraphics[#4]{#1}
\caption{#2} 
\label{#3}
\end{figure}
}

\newtheorem{th.}{Theorem}
\newtheorem{co.}{Corollary}

\begin{document}

\title{
    Approximate complex amplitude encoding algorithm 
    and its application to data classification problems
}


\author{Naoki~Mitsuda}
\affiliation{Sumitomo Mitsui Trust Bank, Ltd., 1-4-1, Marunouchi, Chiyoda-ku, Tokyo 100-8233, Japan}
\affiliation{Quantum Computing Center, Keio University, 3-14-1 Hiyoshi, Kohoku-ku, Yokohama, Kanagawa 223-8522, Japan}

\author{Tatsuhiro~Ichimura}
\affiliation{Sumitomo Mitsui Trust Bank, Ltd., 1-4-1, Marunouchi, Chiyoda-ku, Tokyo 100-8233, Japan}
\affiliation{Quantum Computing Center, Keio University, 3-14-1 Hiyoshi, Kohoku-ku, Yokohama, Kanagawa 223-8522, Japan}

\author{Kouhei~Nakaji}
\affiliation{
Department of Chemistry, University of Toronto, Toronto, Ontario, Canada M5G 1Z8
}
\affiliation{Research Center for Emerging Computing Technologies, National Institute of Advanced Industrial Science and Technology (AIST), 1-1-1 Umezono, Tsukuba, Ibaraki 305-8568, Japan.}
\affiliation{Quantum Computing Center, Keio University, 3-14-1 Hiyoshi, Kohoku-ku, Yokohama, Kanagawa 223-8522, Japan}

\author{Yohichi~Suzuki}
\affiliation{Quantum Computing Center, Keio University, 3-14-1 Hiyoshi, Kohoku-ku, Yokohama, Kanagawa 223-8522, Japan}

\author{Tomoki~Tanaka}
\affiliation{Mitsubishi UFJ Financial Group~Inc.~and MUFG Bank~Ltd., 2-7-1 Marunouchi, Chiyoda-ku, Tokyo 100-8388}
\affiliation{Quantum Computing Center, Keio University, 3-14-1 Hiyoshi, Kohoku-ku, Yokohama, Kanagawa 223-8522, Japan}

\author{Rudy~Raymond}
\affiliation{IBM Quantum, IBM Japan, 
19-21 Nihonbashi Hakozaki-cho, Chuo-ku, Tokyo 103-8510, Japan}
\affiliation{Quantum Computing Center, Keio University, 3-14-1 Hiyoshi, Kohoku-ku, Yokohama, Kanagawa 223-8522, Japan}
\affiliation{Department~of~Computer Science, University~of~Tokyo, 7-3-1 Hongo Bunkyo-ku, Tokyo 113-0033, Japan}

\author{Hiroyuki~Tezuka}
\affiliation{Advanced Research Laboratory, Technology Infrastructure Center, Technology Platform, Sony Group Corporation, 1-7-1 Konan, Minato-ku, Tokyo 108-0075, Japan}
\affiliation{Quantum Computing Center, Keio University, 3-14-1 Hiyoshi, Kohoku-ku, Yokohama, Kanagawa 223-8522, Japan}
\affiliation{Graduate School of Science and Technology, Keio University, 3-14-1 Hiyoshi, Kohoku-ku, Yokohama, Kanagawa 223- 8522, Japan}

\author{Tamiya~Onodera}
\affiliation{IBM Quantum, IBM Japan,
19-21 Nihonbashi Hakozaki-cho, Chuo-ku, Tokyo 103-8510, Japan}
\affiliation{Quantum Computing Center, Keio University, 3-14-1 Hiyoshi, Kohoku-ku, Yokohama, Kanagawa 223-8522, Japan}

\author{Naoki~Yamamoto\thanks{
		e-mail address: \texttt{yamamoto@appi.keio.ac.jp}
	}}
\affiliation{Quantum Computing Center, Keio University, 3-14-1 Hiyoshi, Kohoku-ku, Yokohama, Kanagawa 223-8522, Japan}
\affiliation{Department of Applied Physics and Physico-Informatics, Keio University, Hiyoshi 3-14-1, Kohoku-ku, Yokohama, Kanagawa 223-8522, Japan}


\begin{abstract}
	Quantum computing has a potential to accelerate data processing efficiency, 
	especially in machine learning, by exploiting special features such as the 
	quantum interference. 
	The major challenge in this application is that, in general, the task of loading 
	a classical data vector into a quantum state requires an exponential number of 
	quantum gates. 
	The approximate amplitude encoding (AAE) method, which uses a variational means to approximately load a given real-valued data vector into the amplitude of a quantum state, was recently proposed as a general approach to this problem mainly for near-term devices. 
	However, AAE cannot load a complex-valued data vector, which narrows its application range. 
	In this work, we extend AAE so that it can handle a complex-valued data vector. 
	The key idea is to employ the fidelity distance as a cost function for  
	optimizing a parametrized quantum circuit, where the classical shadow 
	technique is used to efficiently estimate the fidelity and its gradient. 
	We apply this algorithm to realize the complex-valued-kernel binary classifier called the compact Hadamard classifier, and then we present a numerical experiment showing that it enables classification of the Iris dataset and credit card fraud detection. 
\end{abstract}

\maketitle


\section{Introduction}
\label{SECTION-introduction}

Quantum computing is expected to execute information processing tasks that 
classical computers cannot perform efficiently. One of the most promising 
domains in which quantum computing has a potential to boost its performance 
is machine learning 
\cite{wittek,QML-Biamonte,SupervisedQML,Dunjko_2018,QML_PRSA,schuld2021quantum}. 
The advantage of quantum computing in machine learning may come from the 
capability to represent and manipulate an exponential amount of classical data 
using quantum interference 
\cite{giovannetti2008quantum,park2019circuit,de2020circuit}. 
Furthermore, quantum computing can exponentially speed up basic linear algebra 
subroutines \cite{harrow2009quantum}, which are at the core of quantum machine 
learning, such as the support vector machine \cite{rebentrost2014quantum} and 
the principal component analysis \cite{lloyd2014quantum}. 
However, these applications require fault-tolerant quantum computers. 

In recent years, some attempts to implement machine learning algorithms on 
shallow quantum circuits have been made, such as the SWAP-test 
classifier \cite{blank2020quantum}, the Hadamard test classifier (HTC) 
\cite{schuld2017implementing}, and the compact Hadamard classifier (CHC) 
\cite{blank_chc}. 
The central idea underlying these classifiers is that inner products in an 
exponentially large Hilbert space can be directly accessed by measurement 
without expensive subroutines 
\cite{schuld2017implementing,blank2020quantum,park2020theory}. 
It should be noted, however, that these classifiers assume that the classical 
data (i.e., the training data and the test data) have been loaded into the 
amplitudes of a quantum state, i.e., \textit{amplitude encoding}.

In general, the number of gates exponentially grows with the number of qubits 
for realizing the amplitude encoding 
\cite{grover2000synthesis,sanders2019black,plesch2011quantum,shende2005synthesis,grover2002creating,mottonen2004quantum,shende2004quantum}, which might be a major 
bottleneck in practical applications of quantum computation. 
In order to address this issue, Ref.~\cite{knakaji_aae2022} proposed an algorithm 
called the \textit{approximate amplitude encoding (AAE)} that generates 
approximated $n$-qubit quantum states using $\mathcal{O}(\mbox{poly}(n))$ 
gates. 
However, AAE is only applicable to the problem of loading a real-valued data 
vector; that is, it cannot load a complex-valued data vector. 
This limitation narrows the scope of AAE application. 
For example, AAE cannot be applied for preparing an initial state of the CHC 
\cite{blank_chc}, because the CHC encodes the data into both the real and 
imaginary parts of the amplitude of the underlying quantum state. 
Aside from the CHC, an efficient complex-valued data encoding method is also 
required for state preparation of wavepacket dynamical simulations in quantum 
chemistry \cite{ollitrault2020nonadiabatic,chan2022grid}. 

In this paper, we extend the AAE method so that it can approximately load 
a complex-valued data vector onto a shallow quantum circuit. 
We refer to this algorithm as \textit{approximate complex amplitude encoding (ACAE)}.
The key idea is to change the cost function from the maximum mean discrepancy (MMD) 
\cite{liu2018differentiable,coyle2020born}, used in AAE, to the fidelity, which can 
capture the difference in complex amplitude between two quantum states unlike the 
MMD-based cost function. 
A notable point is that the \textit{classical shadow} \cite{preskill_shadow} is 
used to efficiently estimate the fidelity or its gradient. 
As a result, ACAE enables embedding the real and imaginary parts of any 
complex-valued data vector into the amplitudes of a quantum state. 
In addition, we provide an algorithm composed of ACAE and the CHC; this algorithm 
realizes a quantum circuit for binary data classification, using fewer gates 
than the original CHC which requires exponentially many gates to prepare the 
exact initial quantum state. 
We then give a proof-of-principle demonstration of this classification algorithm 
for the benchmark Iris data classification problem. 
Moreover, we apply the algorithm to the credit card fraud detection problem, 
which is a key challenge in financial institutions.


\section{Approximate Complex Amplitude Encoding Algorithm}
\label{SECTION-algorithm}

\subsection{Goal of the approximate complex amplitude encoding  algorithm}
\label{SECTION-goal}

Quantum state preparation is an important subroutine in quantum algorithms that 
process classical data. 
Ideally, this subroutine is represented by a state preparation oracle, $U$, that 
encodes an $N$-dimensional complex vector, 
$\boldsymbol{c} = \{c_0, \dots,c_{N-1}\}, c_k \in \mathbb{C}$, to the amplitudes 
of an $n$-qubit state, $\data$: 
\begin{equation}
	\label{EQUATION-target-state}
	\data = U|0\rangle^{\otimes n} = \sum_{k=0}^{N-1}c_k |k\rangle,
\end{equation}
where the input vector is normalized as $\|\boldsymbol{c}\|=1$. 
Note that $n = \lceil {\log_2(N)}\rceil$. 
Hereafter, the state \eqref{EQUATION-target-state} is referred to as the 
\textit{target state}. 
Recall that, in general, a quantum circuit for generating the target state 
requires $O(2^n)$ controlled-NOT (CNOT) gates \cite{grover2000synthesis,sanders2019black,plesch2011quantum,shende2005synthesis,grover2002creating,mottonen2004quantum,shende2004quantum}, which might destroy any 
possible quantum advantage.

The objective of ACAE is to generate a quantum state that approximates the 
target state \eqref{EQUATION-target-state}, using a parametrized quantum 
circuit (PQC) that is represented by the unitary matrix 
$U(\boldsymbol{\theta})$, with $\boldsymbol{\theta}$ being the vector of parameters. 
Hereafter, we refer to the state generated from the PQC as a \textit{model state}, 
i.e., $|\psi(\boldsymbol{\theta})\rangle = U(\boldsymbol{\theta})|0\rangle^{\otimes n}$. 
We train a PQC to approximate the 
target state except for the global phase; hence, ideally, the model state is 
trained to satisfy 
\begin{equation}
\label{trained_state}
    |\psi(\boldsymbol{\theta})\rangle 
    = U(\boldsymbol{\theta})|0\rangle^{\otimes n} = e^{i\alpha}\data,
\end{equation}
where $e^{i\alpha}$ is the global phase.

Note that, if the elements of $\boldsymbol{c} = \{c_0, \dots,c_{N-1}\}$ are all real-numbers, we can use AAE \cite{knakaji_aae2022}, which also trains a PQC to 
generate an approximating state. 
The training is accomplished by minimizing the cost function given by the MMD 
\cite{liu2018differentiable,coyle2020born} between two probability distributions 
corresponding to the target and the model states. 
However, this method is not applicable for the general complex-valued data 
loading, because each element of the probability distribution is given by 
the squared absolute value of the complex amplitude of the corresponding 
state vector and thus AAE cannot distinguish real and complex numbers of the state.

\subsection{The proposed algorithm}
\label{SECTION-acae}

\subsubsection{Cost function}

In order to execute a complex-valued data loading, it is necessary to introduce 
a measure that reflects the difference between two quantum states with 
complex-valued amplitudes, which cannot be captured by the MMD-based cost function. 
Here we employ the fidelity between the model state 
$|\psi(\boldsymbol{\theta})\rangle$ and the target state $|{\rm Data} \rangle$: 

\begin{align}
    f(\boldsymbol{\theta}) 
     = \textrm{Tr}\left(\rho_{\textrm{model}}(\boldsymbol{\theta})
    \rho_{\textrm{target}}\right)
     = |\langle {\rm Data} | \psi(\boldsymbol{\theta})\rangle|^2, 
\label{eq:fidelity}
\end{align}
where $\rho_{\textrm{model}}(\boldsymbol{\theta})=
|\psi(\boldsymbol{\theta})\rangle \langle\psi(\boldsymbol{\theta})|$ and 
$\rho_{\textrm{target}}=|{\rm Data}\rangle \langle {\rm Data}|$. 
Although, in general, the fidelity can be estimated using the quantum state 
tomography \cite{d2003quantum}, it is highly resource-intensive because this 
procedure requires accurate expectation values for a set of observables whose 
size grows exponentially with respect to the number of qubits. 
For this reason, we employ the classical shadow technique to estimate the 
fidelity.

\subsubsection{Fidelity estimation by classical shadow}

Classical shadow \cite{preskill_shadow} is a method for constructing 
a classical representation that approximates a quantum state using much 
fewer measurements than the case of state tomography. 
The general goal is to predict the expectation values ${o_j}$ for 
a set of $L$ observables, ${O_j}$:
\begin{equation}
\label{observable}
    o_j(\rho)=\textrm{Tr}\left(O_j\rho\right), \: 1\leq j \leq L,
\end{equation}
where $\rho$ is the underlying density matrix. 
The procedure for constructing the predictor is described below.

First, $\rho$ is transformed by the unitary operator $U$ taken from 
the set of random unitaries $\mathcal{U}$ as 
$\rho \rightarrow U\rho U^{\dag}$, and then each qubit is measured 
in the computational basis. 
For a measurement outcome, ${|\hat{b}\rangle}$, the reverse operation 
$U^{\dag}|\hat{b}\rangle\langle \hat{b}|U$ is calculated and stored 
in a classical memory. 
The averaging operation on $U^{\dag}|\hat{b}\rangle\langle \hat{b}|U$ with 
respect to $U \in \mathcal{U}$ is regarded as a quantum channel on $\rho$;
\begin{equation}
\mathbb{E} \left[
 U^{\dagger} |\hat{b}\rangle\!\langle\hat{b}| U\right]
 = \mathcal{M}(\rho),
\end{equation}
which implies 
\begin{equation}
\label{eq:snapshot}
\rho = \mathbb{E} \left[ \mathcal{M}^{-1} \left(U^\dagger
    |\hat{b} \rangle \! \langle \hat{b}| U \right) \right]. 
\end{equation}
The quantum channel $\mathcal{M}$ depends on the ensemble of unitary 
transformations $\mathcal{U}$. 
Equation \eqref{eq:snapshot} gives us a procedure for constructing an 
approximator for $\rho$. 
That is, if the above measurement plus reverse operation is performed 
$N_{\textrm{shot}}$ times for different $U \in \mathcal{U}$, then we 
obtain an array of $N_{\textrm{shot}}$-independent classical 
snapshots of $\rho$:
\begin{equation}
\label{eq:classical_shadow}
    \begin{split}
        \mathsf{S}(\rho ; N_\textrm{shot}) = \left\{ \hat{\rho}_1 = \mathcal{M}^{-1}\left( U_1^\dagger \ket{\hat{b}_1}\! \bra{\hat{b}_1} U_1 \right), \ldots,\right. \\
        \left. \hat{\rho}_{N_\textrm{shot}} = \mathcal{M}^{-1}\left( U_{N_\textrm{shot}}^\dagger \ket{\hat{b}_{N_\textrm{shot}}} \! \bra{\hat{b}_{N_\textrm{shot}}} U_{N_\textrm{shot}} \right) \right\}. 
    \end{split}
\end{equation}
This array is called the classical shadow of $\rho$. 
Once a classical shadow (\ref{eq:classical_shadow}) is obtained, an estimator 
of $\hat{o}_j$ can be calculated as 
\begin{equation}
\label{eq:estimator}
    \hat{o}_j(\rho)=\frac{1}{N_\textrm{shot}}\sum_{i=1}^{N_\textrm{shot}} \textrm{Tr}\left(O_j\hat{\rho}_i\right),
\end{equation}
where each $\hat{\rho}_i$ is the classical snapshot in 
$\mathsf{S}(\rho ; N_\textrm{shot})$. 
Although Reference~\cite{preskill_shadow} proposed to use the median-of-means 
estimator, we employ the empirical mean for simplicity of the implementation. 
Ref.~\cite{preskill_shadow} proved that this protocol has the following 
sampling complexity.

{\bf Theorem} \cite{preskill_shadow}. 
{\it Classical shadows of size $N_\textrm{shot}$ suffice to predict $L$ arbitrary linear target functions $\mathrm{Tr}(O_1 \rho),\ldots,\mathrm{Tr}(O_L \rho)$ up to additive error $\epsilon$ given that
    \begin{equation}
    \label{Nshot}
        N_{\rm{shot}} \geq \mathcal{O}\left( \frac{\log(L)}{\epsilon^2}\max_j \|O_j\|^2_{\rm{shadow}} \right).
    \end{equation}
}

The definition of the shadow norm $\|O_{j}\|_{\mathrm{shadow}}$ depends on 
the ensemble $\mathcal{U}$. 
Two different ensembles can be considered for selecting the random 
unitaries $U$:

$\bullet$ random Clifford measurements, $U$ belongs to the $n$-

~~ qubit Clifford group; and 

$\bullet$ random Pauli measurements, each $U$ is a tensor 

~~ product of single-qubit operations.
\\
For the random Clifford measurements, 
$\|O \|_{\mathrm{shadow}}^2$ is closely related to the Hilbert-Schmidt 
norm $\mathrm{Tr}(O^2)$.
As a result, a large collection of (global) observables with bounded 
Hilbert-Schmidt norm can be predicted efficiently.
For the random Pauli measurements, on the other hand, the shadow norm 
scales exponentially in the locality of the observable; 
note that, for certain cases of the random Pauli measurements, we can use 
the decision-diagram-based classical shadow to have an efficient measurement 
scheme \cite{hillmich2021decision}.

The above classical shadow technique can be directly applied to the problem 
of estimating the fidelity $f(\boldsymbol{\theta})$ given in Eq.~\eqref{eq:fidelity}. 
Actually this corresponds to $L=1$, $O=O_1= \rho_{\textrm{target}}$, and 
$\rho=\rho_{\rm model}(\boldsymbol{\theta})$ in Eq.~(\ref{observable}); 
then, from Eq.~\eqref{eq:estimator}, we have the estimate of $f(\boldsymbol{\theta})$ 
as 
\begin{equation}
\label{eq:fidelity esti}
     \hat{f}(\boldsymbol{\theta}) 
        = \frac{1}{N_\textrm{shot}}\sum_{i=1}^{N_\textrm{shot}} \textrm{Tr}\left(O\hat{\rho}_i(\boldsymbol{\theta})\right), 
\end{equation}
where $\hat{\rho}_i(\boldsymbol{\theta})$ is the classical snapshot of 
$\rho_{\rm model}(\boldsymbol{\theta})$. 
In this case, the random Clifford measurements should be selected, because  
the shadow norm is given by $\mathrm{Tr}(\rho_{\textrm{target}}^2)=1$; 
also, $N_{\rm{shot}}$ becomes independent of system size, because the 
$\max_j \|O_j\|^2_{\rm{shadow}}$ term now becomes constant. 
For the random Clifford measurements, Reference~\cite{preskill_shadow} shows 
that the inverted quantum channel $\mathcal{M}^{-1}$ is given by 
\begin{equation}
    \mathcal{M}^{-1}(\rho) = (2^n+1)\rho-I. 
\end{equation}
Lastly note that $\mathcal{O}\left(n^2/{\log_2(n)}\right)$ entangling gates 
are needed to do sampling from the $n$-qubit Clifford unitaries 
\cite{aaronson2004improved,patel2008optimal}, which is a practical drawback.

\subsubsection{Optimization of $U(\boldsymbol{\theta})$}
\label{subsec:optimization}

\fig{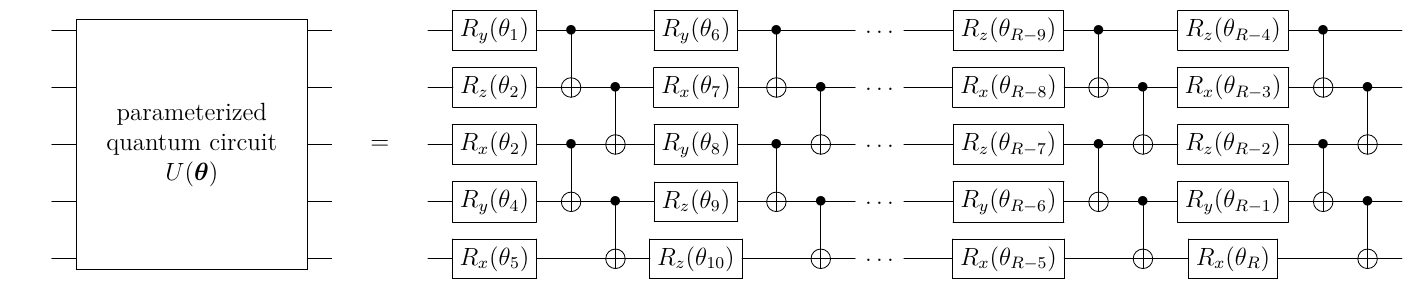}{
Example of the structure of the hardware efficient ansatz $U(\boldsymbol{\theta})$. 
Each layer is composed of the set of parametrized single-qubit rotational gates 
$R_x(\theta_r)=\exp{(-i\theta_r\sigma_x/2)}, R_y(\theta_r)=\exp{(-i\theta_r\sigma_y/2)}${, or} 
 $ R_z(\theta_r)=\exp{(-i\theta_r\sigma_z/2)}$. 
We randomly initialize all axes of each rotating gate (i.e., $X,Y$, or $Z$) 
and $\theta_r$ at the beginning of each training.}
{FIGURE-pqc}{width=450pt}

\fig{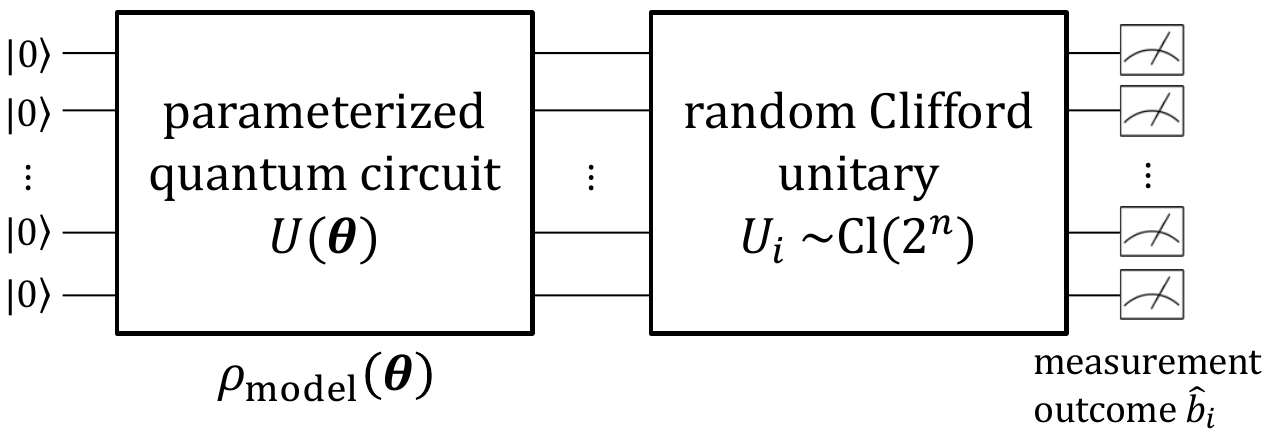}{
Configuration of the $n$-qubit quantum circuit for training PQC 
$U(\boldsymbol{\theta})$. 
The PQC is followed by a unitary, $U_i$, which is selected from the random 
Clifford unitaries with the size $2^n$. 
Measurement is carried out to obtain the outcome $\hat{b}_i$, i.e., a bit 
string of length $n$.}
{FIGURE-training-circuit}
{width=330pt}

Here we describe the training method of the PQC for ACAE. 
The PQC $U(\boldsymbol{\theta})$ consists of $n$ qubits and $l$ layers; thus 
$U(\boldsymbol{\theta})$ contains $\mathcal{O}(ln)$ gates. 
The number of layers $l$ is of the order from $\mathcal{O}(1)$ to 
$\mathcal{O}(\mbox{poly}(n))$. 
In this paper we take the PQC composed of single-qubit parametrized rotational 
gates $R_x(\theta_r)=\exp{(-i\theta_r\sigma_x/2)}$, 
$R_y(\theta_r)=\exp{(-i\theta_r\sigma_y/2)}$, and 
$R_z(\theta_r)=\exp{(-i\theta_r\sigma_z/2)}$ together with CNOT gates; 
here $\theta_r$ is the $r$-th element of $\boldsymbol{\theta}$ and 
$\sigma_{x}, \sigma_y$, and $\sigma_z$ are the Pauli $X, Y$, and $Z$ operators, respectively. 
We take the so-called hardware efficient ansatz \cite{kandala2017}; an example of 
the structure is shown in Fig.~\ref{FIGURE-pqc}. 
This PQC is followed by a random Clifford unitary $U_i$ as 
shown in Fig.~\ref{FIGURE-training-circuit}. 
The output of the circuit is measured $N_{\rm{shot}}$ times in the computational 
basis, with changing the random Clifford unitary $U_i$ and obtaining the outcome 
$\hat{b}_i$ in each trial. 
The above procedure provides us with the classical snapshots of 
$\rho_{\rm{model}}$. 
Note that each $\ket{\hat{b_i}}$ and $U_i$ can be stored efficiently in 
a classical memory using the stabilizer formalism \cite{aaronson2004improved}.

Our goal is to find the best $\boldsymbol{\theta}$ that maximizes the fidelity 
$f(\boldsymbol{\theta})$. 
Now we use the classical shadow to have the estimate 
$\hat{f}(\boldsymbol{\theta})$ given in Eq.~\eqref{eq:fidelity esti}, 
which can be further calculated as
\begin{flalign*}
    \hat{f}(\boldsymbol{\theta})&= \frac{1}{N_\textrm{shot}}\sum_{i=1}^{N_\textrm{shot}} \textrm{Tr}\left(O\hat{\rho}_i(\boldsymbol{\theta})\right)&\\
    &= \frac{1}{N_\textrm{shot}}\sum_{i=1}^{N_\textrm{shot}} \textrm{Tr}\left[O\mathcal{M}^{-1}\left(U^{\dagger}_i\ket{\hat{b}_i}\! \bra{\hat{b}_i}U_i\right)\right]&\\
    &= \frac{1}{N_\textrm{shot}}\sum_{i=1}^{N_\textrm{shot}} \textrm{Tr}\left[O\left\{(2^n+1)U^{\dagger}_i\ket{\hat{b}_i}\! \bra{\hat{b}_i}U_i-I\right\}\right]&\\
    &= \frac{1}{N_\textrm{shot}}\sum_{i=1}^{N_\textrm{shot}}(2^n+1)\bra{\hat{b}_i}U_{i}OU^{\dagger}_i\ket{\hat{b}_i}-1.
\end{flalign*}
Let us consider the run time for evaluating the above 
$\hat{f}(\boldsymbol{\theta})$ with a classical computer. 
First, each $\bra{\hat{b}_i}U_{i}OU^{\dagger}_i\ket{\hat{b}_i}$ is calculated as 
follows: 
\begin{flalign}
    \bra{\hat{b}_i}U_{i}OU^{\dagger}_i\ket{\hat{b}_i}
    &=\bra{\hat{b}_i}U_{i}\ket{{\rm Data}}\bra{{\rm Data}}U^{\dagger}_i\ket{\hat{b}_i} 
    \nonumber \\
    &=\left|\bra{\hat{b}_i}U_i\ket{{\rm Data}}\right|^2& 
    \nonumber \\
    &=\left|\sum_{k=0}^{N-1}c_k\bra{\hat{b}_i}U_i\ket{k}\right|^2.
\label{<b|U|k>}
\end{flalign}
The Gottesman-Knill theorem \cite{gottesman1998heisenberg} allows for evaluation 
of $\bra{\hat{b}_i}U_i\ket{k}$ in $\mathcal{O}(n^2)$ time, because 
$(\ket{\hat{b}_i},\ket{k})$ and $U_i$ are stabilizer states and a Clifford 
operator, respectively. 
Note that the summation in Eq.~(\ref{<b|U|k>}) requires 
$\mathcal{O}(N)=\mathcal{O}(2^n)$ computations, which means that the required 
run time in the training process of the PQC scales exponentially with the number 
of qubits. 
However, this is {\it classical} computation, which should become exponential 
as long as we would like to process a general exponential-size classical data. 
Rather, the advantage of ACAE is in the depth of the PQC, which operates only 
$\mathcal{O}(n \, \mathrm{poly}(n))$ gates instead of $\mathcal{O}(2^n)$, to 
achieve $\mathcal{O}(2^n)$ data encoding.

To maximize the fidelity estimate $\hat{f}(\boldsymbol{\theta})$ {[}i.e., 
to minimize the $-\hat{f}(\boldsymbol{\theta})$){]}, we take the standard gradient 
descent algorithm. 
The gradients of $\hat{f}(\boldsymbol{\theta})$ with respect to ${\theta}_r$ can 
be computed by using the parameter shift rule \cite{Crooks2019} as
\begin{equation}
    \frac{\partial \hat{f}(\boldsymbol{\theta})}{\partial \theta_r} = \hat{f}_{{\theta}_r}^{+} - \hat{f}_{{\theta}_r}^{-},
\end{equation}
where
\begin{equation}
    \hat{f}_{\theta_r}^{\pm}=\hat{f}\left(\theta_1, \cdots, \theta_{r-1}, \theta_r\pm\pi/2, \theta_{r+1}, \cdots, \theta_R\right).
\end{equation}
$R$ denotes the number of the parameters, which can be written as $R = ln$ 
(recall that $l$ is the number of layers of PQC). 
That is, the gradient can also be effectively estimated using the classical shadow. 
This maximization procedure will ideally bring us the optimal parameter set 
$\boldsymbol{\theta}^{*}$ and unitary $U(\boldsymbol{\theta}^{*})$ that generates 
a state approximating the target state (\ref{trained_state}).


\section{Application To Compact Hadamard Classifier}

This section first reviews the method of compact amplitude encoding and 
the compact Hadamard classifier (CHC). 
Then we describe how to apply ACAE to implement the CHC.

\subsection{Compact amplitude encoding}
\label{SECTION-cae}

This method encodes two real-valued data vectors into real and imaginary parts 
of the amplitude of a single quantum state. 
More specifically, given two $N$-dimensional real-valued vectors 
$\boldsymbol{x}^{+}_j= (x_{0j}^{+}, \ldots, x_{(N-1)j}^{+})^T$ and 
$\boldsymbol{x}^{-}_j= (x_{0j}^{-}, \ldots, x_{(N-1)j}^{-})^T$, the compact 
amplitude encoder prepares the following quantum state: 
\begin{equation}
\label{eq:CAE}
    \ket{\boldsymbol{x}_j} := \sum_{l=0}^{N-1} (x_{lj}^{+}+ix^{-}_{lj}) \ket{l},
\end{equation}
where
\begin{equation}
\label{eq:normalization}
    \|\boldsymbol{x}_{j}^{+}\|^2+\|\boldsymbol{x}^{-}_{j}\|^2 = 1
\end{equation}
is assumed to satisfy the normalization condition. 
For simplicity, here we assume $\|\boldsymbol{x}_{j}^{\pm}\|=1/\sqrt{2}$ 
without loss of generality. 
In addition, we define $\ket{\boldsymbol{x}^{\pm}_j}$ as
\begin{equation}
    |\boldsymbol{x}_j^{\pm}\rangle 
    := \frac{1}{\|\boldsymbol{x}_j^{\pm}\|}\sum_{l=0}^{N-1} x_{lj}^{\pm} \ket{l}
    = \sqrt{2}\sum_{l=0}^{N-1} x_{lj}^{\pm} \ket{l}.
\end{equation}

\subsection{Compact Hadamard classifier}
\label{SECTION-chc}

Here we give a quick review about CHC \cite{blank_chc}. 
Suppose the following training data set $\mathcal{D}$ is given as
\begin{equation*}
    \mathcal{D}=\left\{\left(\boldsymbol{x}_0,y_{0}\right), \ldots,\left(\boldsymbol{x}_{M-1},y_{M-1}\right)\right\}.
\end{equation*}
All inputs $\{\boldsymbol{x}_j\}$ are $N$-dimensional real-valued vectors, 
and each $y_j$ takes either $+1$ or $-1$. 
The goal of the CHC is to predict the label $\Tilde{y}$ for a test datum   
$\Tilde{\boldsymbol{x}}$, which is also an $N$-dimensional real-valued vector. 
For simplicity, we assume that the number of training data with label $+1$, 
denoted by $M^{+}$, is equal to the number of training data with label $-1$, 
denoted by $M^{-}$; i.e., {$M^{+}=M^{-}=M/2$,} where $M$ is an even number. 
In particular, we sort the training data set so that 
\begin{equation*}
\boldsymbol{x}_j=
\begin{cases}
\boldsymbol{x}_j^{+}&\text{($0 \leq j\leq M/2-1$),}\\
\boldsymbol{x}_{j-M/2}^{-}&\text{($M/2 \leq j\leq M-1$),}\\
\end{cases}
\end{equation*}
and 
\begin{equation*}
y_j=
\begin{cases}
+1&\text{($0 \leq j\leq M/2-1$),}\\
-1&\text{($M/2 \leq j\leq M-1$)}.
\end{cases}
\end{equation*}
Note that the CHC can also be applied to imbalanced training data sets as we see later.

Assuming the existence of the compact amplitude encoder 
$U_{\mathrm{CAE}}(\boldsymbol{x}_j)$ that encodes the two training data 
vectors $\boldsymbol{x}^{\pm}_j$ into a single quantum state \eqref{eq:CAE} 
and the encoder $U_{\mathrm{AE}}(\boldsymbol{\tilde{x}})$ that encodes 
a test data vector $\tilde{\boldsymbol{x}}$ into the quantum state 
$\ket{\tilde{\boldsymbol{x}}}$, the following quantum state can be generated:
\begin{align}
    \ket{\psi_{\mathrm{init}}}
    &=U_{\mathrm{AE}}(\boldsymbol{\tilde{x}})U_{\mathrm{CAE}}(\boldsymbol{x}_j)R^{(\rm{A})}_zH^{(\rm{A})}U^{(\rm{J})}_w(b)\ket{0}_{\rm{A}}\ket{0}_{\rm{D}}\ket{0}_{\rm{J}},\nonumber\\
    &=U_{\mathrm{AE}}(\boldsymbol{\tilde{x}})U_{\mathrm{CAE}}(\boldsymbol{x}_j)\sum_{j=0}^{\frac{M}{2}-1} \sqrt{b_j}\frac{\ket{0}_{\rm{A}}+{e^{-i\phi}}\ket{1}_{\rm{A}}}{\sqrt{2}}\ket{0}_{\rm{D}}\ket{j}_{\rm{J}},\nonumber\\
    &=\frac{1}{\sqrt{2}} \sum_{j=0}^{\frac{M}{2}-1} \sqrt{b_j}\left(\ket{0}_{\rm{A}}\ket{\boldsymbol{x}_j}_{\rm{D}} + e^{-i\phi}\ket{1}_{\rm{A}}  \ket{\tilde{\boldsymbol{x}}}_{\rm{D}} \right)\ket{j}_{\rm{J}}, 
\label{eq:initial_compact}
\end{align}
where $\phi$ is the relative phase. 
Also, ${b_j}$ is the set of weights satisfying $\sum_{j=0}^{M/2-1} b_j = 1$; 
in this work, we follow {Reference}~\cite{blank_chc} to choose the uniform 
weighting $b_j = 2/M$. 
The labels \rm{A}, \rm{J}, and \rm{D} mean the ancilla qubit, the qubits for 
data numbering, and the qubits for data encoding, respectively 
(the subscript is omitted unless misunderstood).
The number of qubits required to prepare this state is $n+m+1$, where 
$n = \lceil {\log_2(N)}\rceil$ and $m = \lceil {\log_2(M/2)}\rceil$. 
Then, we operate the single-qubit Hadamard gate on the ancilla qubit and get 
\begin{align}
\label{eq:final_compact}
    	\ket{\psi_f} := H^{(\rm{A})}\ket{\psi_{\mathrm{init}}} 
    	= & \frac{1}{2} \sum_{j=0}^{\frac{M}{2}-1} \sqrt{b_j}\big{[} \ket{0}_{\rm{A}}(\ket{\boldsymbol{x}_j}_{\rm{D}} +e^{-i\phi}\ket{\tilde{\boldsymbol{x}}}_{\rm{D}})
    	\nonumber \\
    	& +\ket{1}_{\rm{A}} (\ket{\boldsymbol{x}_j}_{\rm{D}}-e^{-i\phi}\ket{\tilde{\boldsymbol{x}}}_{\rm{D}}) \big{]}\ket{j}_{\rm{J}}.
\end{align}
Finally, we measure the ancilla qubit in the computational basis. 
The entire quantum circuit is illustrated in Fig.~\ref{FIGURE-chc-circuit1}.

\fig{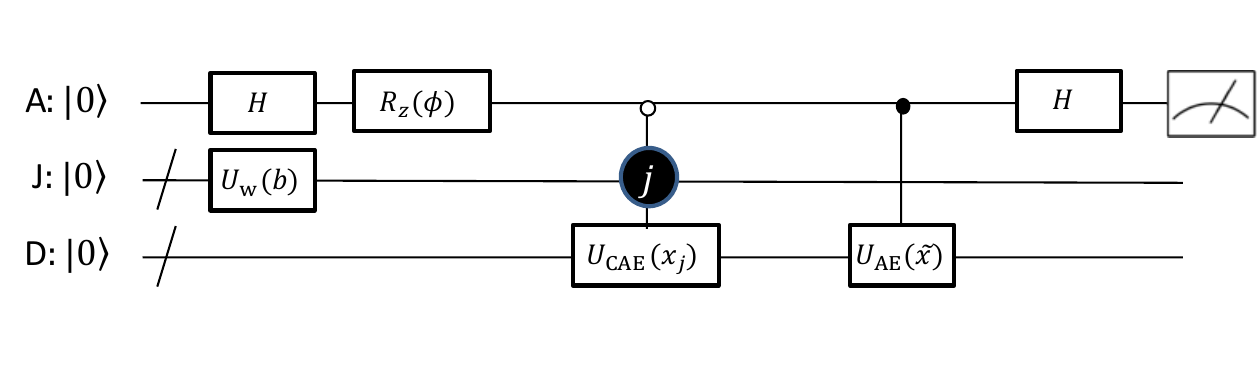}{
Configuration of the circuit for the compact Hadamard classifier. 
The labels \rm{A}, \rm{J}, and \rm{D} on the left side mean the ancilla qubit, 
the qubits for data numbering, and the qubits for data encoding, respectively. 
Slash symbols on the \rm{J} and \rm{D} lines indicate that the line is composed 
of multiple qubits. 
The number of qubits of the \rm{J} and \rm{D} lines are $m = \lceil {\log_2(M/2)}\rceil$ and $n = \lceil {\log_2(N)}\rceil$, 
where $M$ and $N$ represent the number of training data and the dimension of 
the data vector, respectively. 
$U_{\mathrm{w}}(b)$ is an operator that weights each $\ket{j}$ by ${b_j}$.
$U_{\mathrm{CAE}}(\boldsymbol{x}_j)$ and $U_{\mathrm{AE}}(\tilde{\boldsymbol{x}})$ 
are the unitary operators that encode the training data $\boldsymbol{x}^{\pm}_j$ 
and the test data $\tilde{\boldsymbol{x}}$ into $\ket{\boldsymbol{x}_j}_{\rm{D}}$ and 
$\ket{\tilde{\boldsymbol{x}}}_{\rm{D}}$ in Eq.~\eqref{eq:initial_compact}, respectively.
The white and black dots connected to $U_{\mathrm{CAE}}(\boldsymbol{x}_j)$ and $U_{\mathrm{AE}}(\tilde{\boldsymbol{x}})$ 
indicate that the control action turns on when the variable takes 1 and 0, respectively.
The black circle marked {\rm{J}} indicates that the unitary operation 
$U_{\mathrm{CAE}}(\boldsymbol{x}_j)$ is the uniformly controlled via the {\rm{J}} 
qubits in addition to the {\rm{A}} qubit.}{FIGURE-chc-circuit1}{width=450pt}

Now, the probabilities that the measurement outcome of the ancilla qubit is 
$\ket{0}$ and $\ket{1}$ are given by
\begin{align*}
    \Pr(0) = \frac{1}{2}\sum_{j=0}^{\frac{M}{2}-1} b_j\left(1+\cos(\phi)\text{Re}(\kappa_j)-\sin(\phi)\text{Im}(\kappa_j)\right)
\end{align*}
and $\Pr(1)=1-\Pr(0)$, where 
$\kappa_j=\langle\tilde{\boldsymbol{x}}|\boldsymbol{x}_j\rangle$. 
Therefore, the expectation value of the Pauli $Z$ operator measured on the 
ancilla qubit, denoted as $\sigma_z^{(\rm{A})}$, is
\begin{equation*}
    \langle \sigma_z^{(\rm{A})}\rangle = \sum_{j=0}^{M/2-1}b_j\left(\cos(\phi)\text{Re}(\kappa_j)-\sin(\phi)\text{Im}(\kappa_j)\right).
\end{equation*}
If we set $\phi=\pi/4$, this becomes
\begin{equation}
\label{eq:sigmaz1_chc}
    \langle \sigma_z^{(\rm{A})}\rangle = \frac{1}{2}\sum_{j=0}^{\frac{M}{2}-1}b_j\left(\langle\tilde{\boldsymbol{x}}|\boldsymbol{x}_j^{+}\rangle - \langle\tilde{\boldsymbol{x}}|\boldsymbol{x}_j^{-}\rangle\right).
\end{equation}
Note that, when the number of training data vectors in the two classes are not 
equal (i.e., $M^{+}\neq M^{-}$), this difference can be compensated by 
tuning $\phi$ to satisfy $\tan(\phi)=M_{-}/M_{+}$. 
We now end up with the final form of Eq.~\eqref{eq:sigmaz1_chc} as 
\begin{align}
\label{eq:chtc_final}
    \langle \sigma_z^{(\rm{A})}\rangle = \frac{1}{2}\sum_{j=0}^{M-1}b'_jy_j\langle \tilde{\boldsymbol{x}}|\boldsymbol{x}_j\rangle,
\end{align}
where $b'_{j}$ is defined as 
\begin{align*}
   b'_{j} = b'_{j+M/2} = b_j, ~~
   \sum_{j=0}^{M-1}b'_j = 2.
\end{align*}
Clearly Eq.~\eqref{eq:chtc_final} has the form of a standard kernel-based 
classifier where $\langle \tilde{\boldsymbol{x}}|\boldsymbol{x}_j\rangle$ 
represents the similarity between the test data $\tilde{\boldsymbol{x}}$ and 
the training data $\boldsymbol{x}_j$. 
Because the right-hand side of Eq.~\eqref{eq:chtc_final} represents the sum 
of the kernel weighted by $b'_jy_j$, the sign of $\langle \sigma_z^{(\rm{A})}\rangle$ 
tells us the class for the test data $\tilde{\boldsymbol{x}}$; 
that is, the CHC predicts the label $\tilde{y}$ of $\tilde{\boldsymbol{x}}$ via 
the following policy: 
\begin{equation}
\label{eq:classifier}
    \tilde{y} = \sgn{\langle \sigma_z^{(\rm{A})}\rangle}
     = \sgn{ \frac{1}{2}\sum_{j=0}^{M-1}b'_jy_j\langle \tilde{\boldsymbol{x}}|\boldsymbol{x}_j \rangle  }. 
\end{equation}
Note that the weights $\{b_j'\}$ can be optimized, like the standard 
kernel-based classifier such as the support vector machine which indeed 
optimizes those weights depending on the training dataset; 
this clearly improves the classification performance of ACAE, which 
will be examined in a future work.

The advantage of quantum kernel-based classifiers over classical classifiers 
is the accessibility to kernel functions. 
Actually, the kernel (or similarity) between test data and training data is 
calculated as the inner product in the feature Hilbert space, which is 
computationally expensive to evaluate via classical means when the feature 
space is large. 
On the other hand, quantum kernel-based classifiers efficiently evaluate 
kernel functions. 
In particular, the CHC can evaluate the sum of all the inner products in the 
$N$-dimensional feature space appearing in the right-hand side of 
Eq.~\eqref{eq:chtc_final}, just by measuring the expected value of 
$\sigma_z^{(\rm{A})}$. 
We also emphasize that the CHC can be realized with compact quantum circuits 
compared with other quantum kernel-based classifiers. 
Actually, thanks to the compact amplitude encoding, 
two qubits can be removed in the CHC formulation compared to the others; 
moreover, the number of operations for encoding the training data set 
$\boldsymbol{x}_j$ is reduced by a factor of 4 compared with the HTC 
\cite{schuld2017implementing}. 
Hence, the CHC can be implemented in a compact quantum circuit in both depth 
and width compared to the other quantum classifiers, meaning that a smaller 
and thus easier-trainable variational circuit may function for the CHC.

\subsection{Implementation of the CHC using ACAE}

\fig{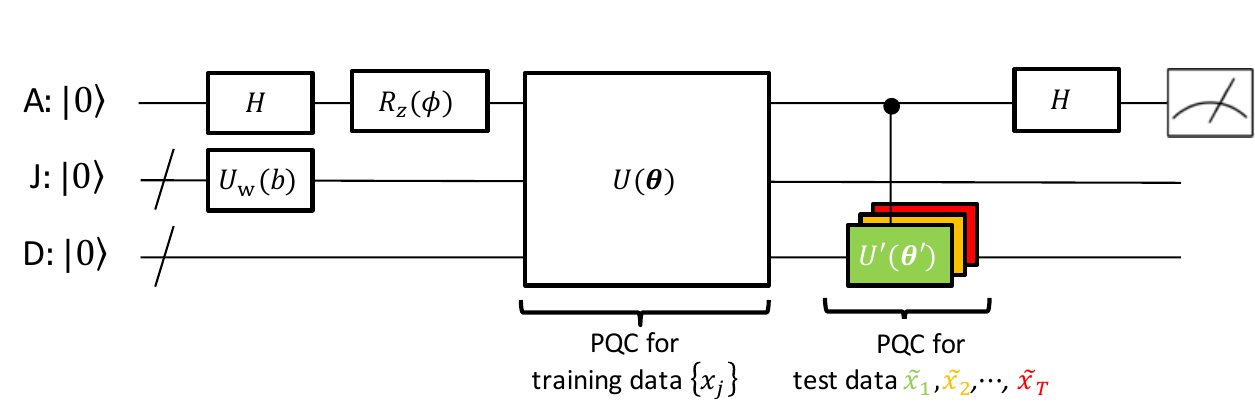}{Configuration of the circuit for the compact Hadamard classifier using ACAE. 
We use PQCs instead of the uniformly controlled $U_{\mathrm{CAE}}(\boldsymbol{x}_j)$ and $U_{\mathrm{AE}}(\tilde{\boldsymbol{x}})$ in Fig.~\ref{FIGURE-chc-circuit1}; 
that is, $U(\boldsymbol{\theta})$ represents a PQC for encoding the training 
data $\{\boldsymbol{x}_j\}$ and thereby approximately generates 
$\ket{\boldsymbol{x}_j}_{\rm{D}}$ in Eq.~(\ref{eq:final_compact}). 
Also, $U'(\boldsymbol{\theta'})$ represents a PQC for encoding the test data 
$\tilde{\boldsymbol{x}}$ and thereby approximately generates 
$\ket{\tilde{\boldsymbol{x}}}_{\rm{D}}$. 
Note that, in practice, we construct $U(\boldsymbol{\theta})$ off-line using 
all the training data; then we learn $U'(\boldsymbol{\theta'})$ for a given test 
datum $\tilde{\boldsymbol{x}}$ for the on-line prediction purpose. 
}{FIGURE-chc-circuit2}{width=450pt}

Although the CHC efficiently realizes a compact quantum classifier, it relies 
on the critical assumption; that is, the quantum state \eqref{eq:CAE} is 
necessarily prepared. 
Recall that, in general, the quantum circuit for generating the state 
\eqref{eq:CAE} requires an exponential number of gates. 
Moreover, to generate the quantum state $\ket{\psi_{\mathrm{init}}}$ in 
Eq.~\eqref{eq:initial_compact}, the uniformly controlled 
gate~\cite{mottonen2004transformation,bergholm2005quantum} shown in 
Fig.~\ref{FIGURE-chc-circuit1} also requires an exponential number of gates.
These requirements may destroy the quantum advantage of the CHC.

The ACAE implements the CHC without using exponentially many gates; that is, 
as shown below, we can approximately generate the quantum state $\ket{\psi_f}$ 
with a constant-depth quantum circuit illustrated in Fig.~\ref{FIGURE-chc-circuit2}. 
First, by applying $H, R_z(\phi)$, and $U_w(b)$ on the initial states, we produce 
the following state:
\begin{equation}
    \ket{\psi_0}:=\frac{1}{\sqrt{2}} \sum_{j=0}^{\frac{M}{2}-1} \sqrt{b_j}\left(\ket{0}_{\rm{A}}\ket{0}_{\rm{D}} + e^{-i\phi}\ket{1}_{\rm{A}}  \ket{0}_{\rm{D}} \right)\ket{j}_{\rm{J}}.
\end{equation} 
Next, we train $U(\boldsymbol{\theta})$ by the algorithm described in Sec. II, 
so that it approximately encodes the training data $\{\boldsymbol{x}_j\}$ to have 
\begin{align}
\label{eq:encoding_only_training_data}
    \ket{\psi_1}&:=U(\boldsymbol{\theta})\ket{\psi_0}\notag\\
    & \approx e^{i\alpha}\frac{1}{\sqrt{2}} \sum_{j=0}^{\frac{M}{2}-1} \sqrt{b_j}\left(\ket{0}_{\rm{A}}\ket{\boldsymbol{x}_j}_{\rm{D}} + e^{-i\phi}\ket{1}_{\rm{A}}  \ket{0}_{\rm{D}}\right)\ket{j}_{\rm{J}},
\end{align}
where $e^{i\alpha}$ is the global phase. 
We then set the controlled $U'(\boldsymbol{\theta'})$ to encode the test data 
$\tilde{\boldsymbol{x}}$ to obtain
\begin{align}
    &(\ket{0}_{\rm{A}}\bra{0}_{\rm{A}}\otimes I + \ket{1}_{\rm{A}}\bra{1}_{\rm{A}}\otimes U'(\boldsymbol{\theta'}))\ket{\psi_1}\notag\\
    &\approx e^{i\alpha}\frac{1}{\sqrt{2}} \sum_{j=0}^{\frac{M}{2}-1} \sqrt{b_j}\left(\ket{0}_{\rm{A}}\ket{\boldsymbol{x}_j}_{\rm{D}} + e^{-i\phi}\ket{1}_{\rm{A}}  \ket{\tilde{\boldsymbol{x}}}_{\rm{D}} \right)\ket{j}_{\rm{J}}\notag\\&=e^{i\alpha}\ket{\psi_{\mathrm{init}}}
\label{eq:psi_i}.
\end{align}
Finally, we operate $H$ on the ancilla qubit to obtain {$\ket{\psi_f}$:} 
\begin{equation}
    H^{(\rm{A})}e^{i\alpha}\ket{\psi_{\mathrm{init}}}=e^{i\alpha}\ket{\psi_f}.
\end{equation} 

Although the global phase $e^{i\alpha}$ is added to $\ket{\psi_f}$, this does 
not affect the probability $\Pr(0)$ and the expectation value 
$\langle \sigma_z^{(\rm{A})}\rangle$. 
Recall that, $U(\boldsymbol{\theta})$ and $U'(\boldsymbol{\theta'})$ consist of 
$(n+m+1)$ qubits with $\mathcal{O}(1)\sim \mathcal{O}(\mbox{poly}(n+m+1))$ layers 
and $n$ qubits with $\mathcal{O}(1)\sim \mathcal{O}(\mbox{poly}(n))$ layers, 
respectively.
Therefore, we can implement the approximated CHC without using exponentially 
many gates.

Before moving forward, we give some remarks, regarding the design of 
$U'(\boldsymbol{\theta'})$. 
First, we assume that $U'(\boldsymbol{\theta'})$ consists of only $R_y$ gates 
and CNOT gates, because $\tilde{\boldsymbol{x}}$ is a real-valued data vector. 
As a result, the unitary process is restricted to 
$U'(\boldsymbol{\theta'})\ket{0}_{\rm{D}} \approx \pm \ket{\tilde{\boldsymbol{x}}}_{\rm{D}}$, 
which is further converted to 
$U'(\boldsymbol{\theta'})\ket{0}_{\rm{D}} \approx \ket{\tilde{\boldsymbol{x}}}_{\rm{D}}$ 
by compensating the phase via the choice of $\phi$ in Eq.~\eqref{eq:psi_i}. 
Next, to implement the controlled $U'(\boldsymbol{\theta'})$, every elementary gate 
contained in $U'(\boldsymbol{\theta'})$ has to be modified to a controlled gate 
via the {\rm{A}} qubit. 
Finally, in practice, we construct $U(\boldsymbol{\theta})$ off-line using 
all the training data; once a test datum $\tilde{\boldsymbol{x}}$ to be classified 
is given to us, we learn $U'(\boldsymbol{\theta'})$ to approximately encode 
$\tilde{\boldsymbol{x}}$ and then construct the CHC to predict the corresponding label 
$\tilde{y}$. 


\section{Demonstration}
\label{SECTION-demo}

\begin{table*}[htb]
    \caption{The data contents embedded in the quantum amplitude of $\ket{\psi_{\mathrm{init}}}$ in {$Iris$ $setosa$} and {$Iris$ $versicolor$} classification problem. The $\rm{A}$, $\rm{J}$, and $\rm{D}$ in the left three columns represent the ancilla qubit, the qubits for data numbering, and the qubits for data encoding, respectively, as in Fig.~\ref{FIGURE-chc-circuit1}. The ``$i$" in the right end column represents an imaginary unit. The amplitudes are normalized to satisfy Eq.~\eqref{eq:normalization}.}
    \label{TABLE-data}
    \centering
    \begin{tabular}{|c|c|c|c||c|}
    \hline
        \rm{A}  & \rm{J} & \rm{D} & \multirow{4}{*}{Basis} & \multirow{4}{*}{Data contents embedded in the amplitude of each basis}  \\
         Ancilla qubit & Qubits for & Qubits for &  & \\
          0 : training data & data numbering  & data enconding &  &  \\
          1 : test data\qquad\qquad & (from 0 to $M/2$) & (from 0 to $M$) &  &  \\ \hline\hline
        \multirow{9}{*}{0} & \multirow{4}{*}{00} & 00 & $\ket{00000}$ & (1st feature of {$I.$ $setosa$}) + (1st feature of {$I.$ $versicolor$})$\times i$  \\ \cline{3-5}
         &  & 01 & $\ket{00001}$ & (2nd feature of {$I.$ $setosa$}) + (2nd feature of {$I.$ $versicolor$})$\times i$  \\ \cline{3-5}
         &  & 10 & $\ket{00010}$ & (3rd feature of {$I.$ $setosa$}) + (3rd feature of {$I.$ $versicolor$})$\times i$  \\ \cline{3-5}
         &  & 11 & $\ket{00011}$ & (4th feature of {$I.$ $setosa$}) + (4th feature of {$I.$ $versicolor$})$\times i$  \\ \cline{2-5}
         & $\vdots$ & $\vdots$ & $\vdots$ & $\vdots$  \\ \cline{2-5}
         & \multirow{4}{*}{11} & 00 & $\ket{01100}$ & (1st feature of {$I.$ $setosa$}) + (1st feature of {$I.$ $versicolor$})$\times i$  \\ \cline{3-5}
         &  & 01 & $\ket{01101}$ & (2nd feature of {$I.$ $setosa$}) + (2nd feature of {$I.$ $versicolor$})$\times i$  \\ \cline{3-5}
         &  & 10 & $\ket{01110}$ & (3rd feature of {$I.$ $setosa$}) + (3rd feature of {$I.$ $versicolor$})$\times i$  \\ \cline{3-5}
         &  & 11 & $\ket{01111}$ & (4th feature of {$I.$ $setosa$}) + (4th feature of {$I.$ $versicolor$})$\times i$  \\ \hline
        \multirow{9}{*}{1} & \multirow{4}{*}{00} & 00 & $\ket{10000}$ & (1st feature of the test data)$\times \exp(-i\phi)$  \\ \cline{3-5}
         &  & 01 & $\ket{10001}$ & (2nd feature of the test data)$\times \exp(-i\phi)$  \\ \cline{3-5}
         &  & 10 & $\ket{10010}$ & (3rd feature of the test data)$\times \exp(-i\phi)$  \\ \cline{3-5}
         &  & 11 & $\ket{10011}$ & (4th feature of the test data)$\times \exp(-i\phi)$  \\ \cline{2-5}
         & $\vdots$ & $\vdots$ & $\vdots$ & $\vdots$  \\ \cline{2-5}
         & \multirow{4}{*}{11} & 00 & $\ket{11100}$ & (1st feature of the test data)$\times \exp(-i\phi)$  \\ \cline{3-5}
         &  & 01 & $\ket{11101}$ & (2nd feature of the test data)$\times \exp(-i\phi)$  \\ \cline{3-5}
         &  & 10 & $\ket{11110}$ & (3rd feature of the test data)$\times \exp(-i\phi)$  \\ \cline{3-5}
         &  & 11 & $\ket{11111}$ & (4th feature of the test data)$\times \exp(-i\phi)$ \\ \hline
    \end{tabular}
\end{table*}

This section gives two numerical demonstrations to show the performance of 
our algorithm composed of ACAE and the CHC. 
First we present an example application to a classification problem for the 
Iris dataset \cite{iris}.
Next we show application to the fraud detection problem using the credit 
card fraud dataset \cite{credit} provided in Kaggle.

\subsection{Iris dataset classification}
\label{SECTION-iris}

The Iris dataset consists of three iris species ({$Iris$ $setosa$, $Iris$ $virginica$, and $Iris$ $versicolor$}) with 50 samples each as well as four features (sepal 
length, sepal width, petal length, and petal width) for each flower. 
Each sample data includes the identification (ID) number, four features, and the species. 
IDs for 1 to 50, 51 to 100, and 101 to 150 represent data for $I.$ $setosa$, $I.$ $virginica$, and $I.$ $versicolor$, respectively.

In this paper, we consider the {$I.$ $setosa$} and {$I.$ $versicolor$} classification 
problem. 
The goal is to create a binary classifier that predicts the correct label 
$\tilde{y}$ (0 : {$I.$ $setosa$}, 1 : {$I.$ $versicolor$}) for the given test data  
$\tilde{\boldsymbol{x}}$ {= (sepal length, sepal width, petal length, petal width).} 
In this demonstration, we employ the first four data of each species as training 
data. 
That is, we use data with IDs 1 to 4 and 51 to 54 as training data for {$I.$ $setosa$} and {$I.$ $versicolor$}, respectively. 
On the other hand, we use data with IDs 5 to 8 and 55 to 58 as test data.

First, we need to prepare the quantum state $\ket{\psi_{\mathrm{init}}}$ given 
in Eq.~(\ref{eq:initial_compact}) using ACAE. 
Since the dimension of the feature vector and the number of training data are $N=4$ 
and $M=8$, the number of required qubits are $n = \lceil {\log_2(N)}\rceil=2$ and 
$m = \lceil {\log_2(M/2)}\rceil=2$, respectively. 
The number of total qubits required for composing the quantum circuit is $n+m+1=5$, 
which means that $\ket{\psi_{\mathrm{init}}}$ has $2^5=32$ amplitudes. 
{Table}~\ref{TABLE-data} shows the data contents embedded in the quantum amplitude 
of each basis.

We encode the training data of the {$I.$ $setosa$} and {$I.$ $versicolor$} into 
the complex amplitude of the ancilla qubit state $\ket{0}$ by using the PQC $U(\boldsymbol{\theta})$. 
Also, we encode the test data into the amplitude of the ancilla qubit state 
$\ket{1}$ by using $U'(\boldsymbol{\theta'})$.
We use the 5-qubit and 12-layer ansatz $U(\boldsymbol{\theta})$ illustrated in 
Fig.~\ref{FIGURE-pqc}.
We randomly initialize all the directions of each rotating gate (i.e., $X, Y$, or $Z$) 
and $\theta_r$ at the beginning of each training. 
As the optimizer, Adam \cite{adam} is used. 
The number of iterations (i.e., the number of the updates of the parameters) is 
set to $400$ for training $U(\boldsymbol{\theta})$.
For each iteration, 1000 classical snapshots are used to estimate the fidelity; 
that is, we set $N_\textrm{shot}=1000$ in Eq.~(\ref{eq:estimator}). 
The learning rate is 0.1 for the first $100$ iterations, 0.01 for the next $100$ 
iterations, 0.005 for the next $100$ iterations, and 0.001 for the last $100$ 
iterations. 
In Fig.~\ref{FIGURE-fidelity_iris}, we show the change of fidelity in the training 
process of $U(\boldsymbol{\theta})$. 
At the end of the training, the fidelity reaches 0.994.

As for the encoding process of test data, we use the 2-qubit and 2-layer 
ansatz $U'(\boldsymbol{\theta'})$ that consists of only $R_y$ gates and CNOT 
gates.
The number of iterations and $N_\textrm{shot}$ are $100$ and $1000$, respectively.
The learning rate is 0.1 for the first $50$ iterations and 0.01 for the next $50$. 
We found that, for each test datum, the fidelity reaches the value bigger than 
0.999. 
It is notable that the number of layers of $U'(\boldsymbol{\theta'})$ and 
the number of iterations for training it are much smaller compared to the 
encoding process for training data. 
This means, once the training data are encoded, it is easy to change the test data.

\fighalf{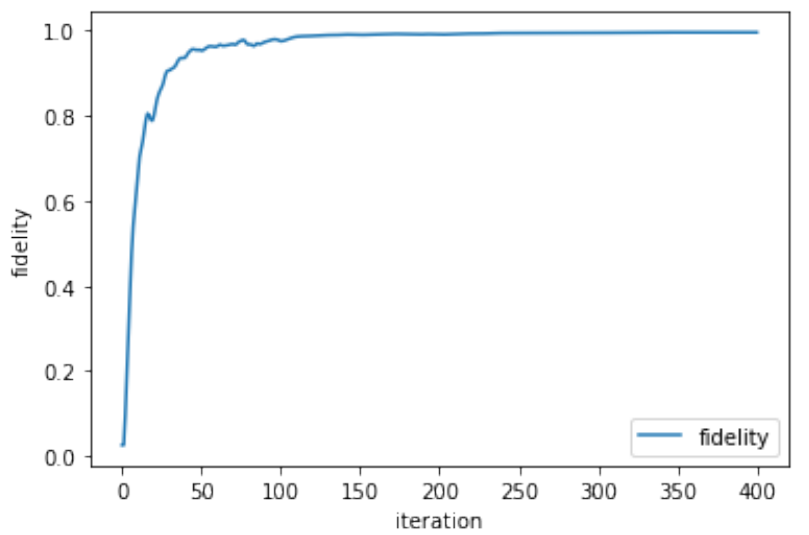}{
The change of the fidelity between the target state and the model state in the training process of $U(\boldsymbol{\theta})$. 
Here, the target state is {Eq.} (\ref{eq:encoding_only_training_data}), and the 
model state means the state that is actually generated by the PQC 
$U(\boldsymbol{\theta})$.}
{FIGURE-fidelity_iris}{width=240pt}

\begin{figure}[htb]
    \centering
    \subfigure[]{
        \includegraphics[width=240pt]{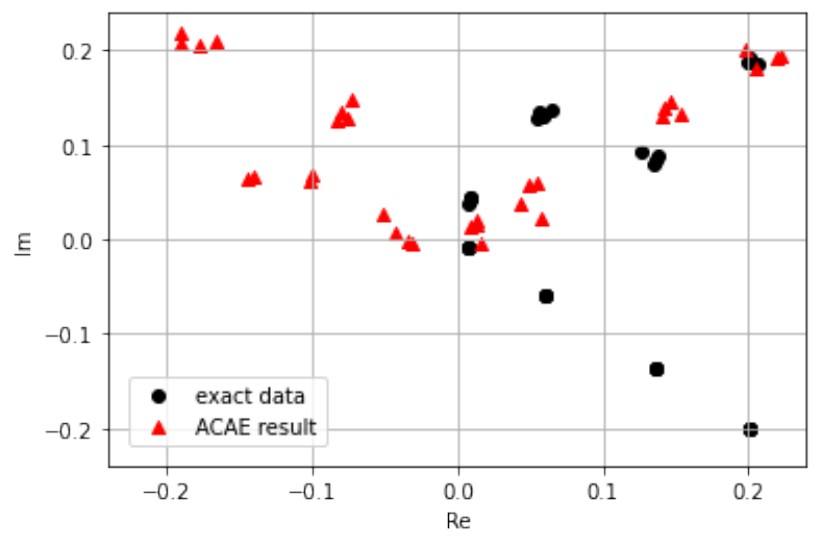}
    }\\
    \subfigure[]{
        \includegraphics[width=240pt]{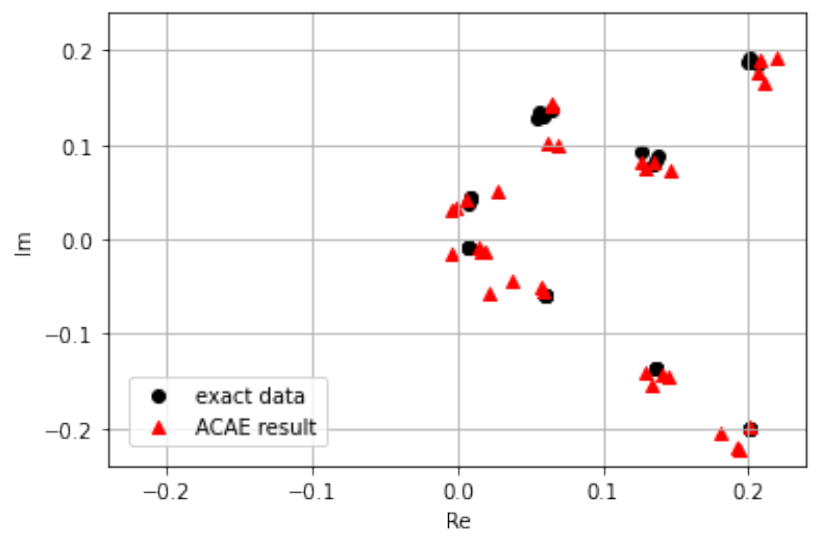}
    }
    \caption{An example of data encoding results. Note that each data value is divided by a constant number to satisfy the normalization condition (\ref{eq:normalization}).
    \\(a) The complex amplitudes of $\ket{\psi_{\mathrm{init}}}$ generated by ACAE (red triangles) and exact data (black dots). Note that red dots contain the influence of the global phase $e^{i\alpha}$.
    \\(b) The result in which the global phase is hypothetically eliminated is shown. The fidelity between the model state (red triangles) and the target state (black dots) is 0.993.} 
    \label{FIGURE-amplitudes_iris}
\end{figure}

As an example of the data encoding results, a set of complex amplitudes generated 
by ACAE is shown in Fig.~\ref{FIGURE-amplitudes_iris}.
In the figure, the value of each complex amplitude is plotted on the complex plane.
The black dots and red triangles represent the exact data and the approximate complex amplitudes embedded by ACAE, respectively.
Note that the complex amplitude embedded by ACAE contains the global phase $e^{i\alpha}$ 
as in Eq.~(\ref{eq:psi_i}).
In order to compare the exact data with the ACAE result visually, the result in 
which the global phase is hypothetically eliminated is also shown. 
Recall that the global phase does not affect the measurement in the remaining 
procedure.

After the state preparation, we operate the Hadamard gate on the ancilla qubit 
and obtain $\ket{\psi_f}$ in Eq.~(\ref{eq:final_compact}).
By measuring the ancilla qubit of this quantum state and obtaining the sign of 
$\langle \sigma_z^{(\rm{A})}\rangle$, we can use Eq.~\eqref{eq:classifier} to predict 
the label $\tilde{y}$ of the test data $\tilde{\boldsymbol{x}}$, i.e., 
$\tilde{y} = \sgn{\langle \sigma_z^{(\rm{A})}\rangle}$.
Classification results are shown in Table \ref{tbl:iris1} for 8 cases in which 
the test data are IDs 5 to 8 and IDs 55 to 58.
In addition to the {$I.$ $setosa$} and {$I.$ $versicolor$} classification problem, we 
carry out the {$I.$ $versicolor$} and {$I.$ $virginica$} classification problems in the 
same way and show the results in {Table}~\ref{tbl:iris2}.
All classification results are correct in {Table}~\ref{tbl:iris1}.
On the other hand, three out of eight classification results are incorrect in 
{Table}~\ref{tbl:iris2}.

Let us discuss the results.
Pairwise relationships in the Iris dataset shows that {$I.$ $setosa$} can be clearly 
separated from the other two varieties, whereas the features of {$I.$ $versicolor$} 
and {$I.$ $virginica$} slightly overlap with each other. 
Therefore, classification of {$I.$ $versicolor$} and {$I.$ $virginica$} {[Table~\ref{tbl:iris2}]} 
is considered more difficult than that of {$I.$ $setosa$} and {$I.$ $versicolor$} 
{[Table~\ref{tbl:iris1}]}. 
This could be the cause of the low accuracy rate of {Table}~\ref{tbl:iris2}.
Note that the error between the exact data and the ACAE data also affects the 
classification accuracy.
If we adjust the number of layers in the PQC and the number of iterations 
in the training process to improve the fidelity between the target state and 
the model state, we will be able to increase the accuracy rate. 
In fact, we have confirmed that the incorrect results 
(\#55, \#56, \#57) turned correct when the classifications 
were performed using the exact data.
{Table}~\ref{tbl:iris3} shows the results where both the training 
and the test data are ideally encoded without errors.

\begin{table}
    \centering
    \caption{The results of the Iris dataset classification problem \label{tbl:iris}}

    \subtable[{$Iris$ $setosa$} versus {$Iris$ $versicolor$}]{
        \begin{tabular}{|c|c|c|c|c|}  \hline
            {Test} data ID & {Class} & $\quad\langle \sigma_z^{(\rm{A})}\rangle\quad$ & $\quad\tilde{y}\quad$ & $\quad${Result}$\quad$\\ \hline \hline
            \#5 &  & 0.0422 & $+$ & Correct\\
            \#6 & {$I.$ $setosa$} & 0.0364 & $+$ & Correct\\ 
            \#7 & ($+$) & 0.0395 & $+$ & Correct\\
            \#8 &  & 0.0374 & $+$ & Correct\\ \hline
            \#55 &  & -0.0317 & $-$ & Correct\\
            \#56 & {$I.$ $versicolor$} & -0.0315 & $-$ & Correct\\ 
            \#57 & ($-$) & -0.0285 & $-$ & Correct\\
            \#58 &  & -0.0239 & $-$ & Correct\\ \hline
        \end{tabular}
        \label{tbl:iris1}
    }

    \subtable[{$Iris$ $versicolor$} versus {$Iris$ $virginica$}]{
        \begin{tabular}{|c|c|c|c|c|}  \hline
            {Test} data ID & {Class} & $\quad\langle \sigma_z^{(\rm{A})}\rangle\quad$ & $\quad\tilde{y}\quad$ & $\quad${Result}$\quad$\\ \hline \hline
            \#55 &  & -0.0001 & $-$ & Incorrect\\
            \#56 & {$I.$ $versicolor$} & -0.0015 & $-$ & Incorrect\\ 
            \#57 & $(+)$ & -0.0020 & $-$ & Incorrect\\
            \#58 &  & 0.0010 & $+$ & Correct\\ \hline
            \#105 &  & -0.0083 & $-$ & Correct\\
            \#106 & {$I.$ $virginica$} & -0.0066 & $-$ & Correct\\ 
            \#107 & $(-)$ & -0.0089 & $-$ & Correct\\
            \#108 &  & -0.0054 & $-$ & 
            Correct\\ \hline
        \end{tabular}
        \label{tbl:iris2}
    }
    
    \subtable[{$Iris$ $versicolor$} versus {$Iris$ $virginica$}(Exact data are used.)]{
        \begin{tabular}{|c|c|c|c|c|}  \hline
            {Test} data ID & {Class} & $\quad\langle \sigma_z^{(\rm{A})}\rangle\quad$ & $\quad\tilde{y}\quad$ & $\quad${Result}$\quad$\\ \hline \hline
            \#55 &  & 0.0322 & $+$ & Correct\\
            \#56 & {$I.$ $versicolor$} & 0.0013 & $+$ & Correct\\ 
            \#57 & $(+)$ & 0.0230 & $+$ & Correct\\
            \#58 &  & 0.0534 & $+$ & Correct\\ \hline
            \#105 &  & -0.0392 & $-$ & Correct\\
            \#106 & {$I.$ $virginica$} & -0.0269 & $-$ & Correct\\ 
            \#107 & $(-)$ & -0.0430 & $-$ & Correct\\
            \#108 &  & -0.0194 & $-$ & 
            Correct\\ \hline
        \end{tabular}
        \label{tbl:iris3}
    }
\end{table}

\subsection{Credit card fraud detection}
\label{SECTION-fraud_detection}

Nowadays, credit card fraud is a social problem in terms of customer protection, financial crime prevention, and avoiding negative impacts on corporate finances.
The losses that arise from credit card fraud are a serious problem for financial institutions; according to the Nilson Report \cite{nilson}, credit card fraud 
losses are expected to reach \$49.3 billion by 2030. 
Banks and credit card companies that pay for fraud will be hit hard by these 
rising costs. 
With digital crime and online fraud on the rise, it is more important than ever for financial institutions to prevent credit card fraud through advanced technology and strong security measures.
To detect fraudulent use, financial institutions use human judgment to determine fraud based on information such as cardholder attributes, past transactions, and product delivery address information; however, this method requires human resources and costs.
Although attempts to detect fraud by machine learning based on features extracted from credit card transaction data also have been attracting attention in recent years, there are disadvantages such as more time and epochs to converge for a stable prediction, excessive training and so on.
Quantum machine learning has the potential to solve these challenges, and its application to credit card fraud detection deserves exploring.

In this subsection, we demonstrate credit card fraud detection as another practical application of the CHC with ACAE.
The goal is to encode credit card transaction data into a quantum state as training data and classify whether a given transaction datum $\tilde{\boldsymbol{x}}$ is a normal ($\tilde{y}=+1$) or fraudulent transaction ($\tilde{y}=-1$).

In this demonstration we use the credit card fraud detection dataset \cite{credit} provided by Kaggle.
The dataset contains credit card transactions made by European cardholders in September 2013.
The dataset has 284,807 transactions which include 492 fraudulent transactions.
Note that, since it is difficult to encode all transaction data into a quantum state, for proof-of-concept testing, we take 4 normal transaction data\footnote{We take the top 4 normal data IDs in ascending order, specifically, \#1, \#2, \#3, and \#4.} and 4 fraudulent transaction data\footnote{We take the top 4 fraudulent data IDs in ascending order, specifically, \#524, \#624, \#4921, and \#6109. } from this dataset as training data and perform classification tests to determine whether the given test data are normal or fraudulent.
Each set of data consists of \textit{Times}, \textit{Amount,} and 28 different features ($V1,V2,\dots,V28$) transformed by principal component analysis.
We use 4 features ($V1, V2, V3, V4$) out of the 28 features for classification, which means that the dimension of the feature vector and the number of training data are $N=4$ and $M=8$, respectively, and the number of total qubits required for composing the quantum circuit is $n+m+1=5$.
As in the previous subsection, we embed the training data of normal transactions and fraudulent transactions into the real and imaginary parts of the complex amplitude of the ancilla qubit state $\ket{0}$ and embed the test data into the complex amplitude of the ancilla qubit state $\ket{1}$.

After the state preparation, we operate the Hadamard gate on the ancilla qubit and make measurements to obtain {$\langle \sigma_z^{(\rm{A})}\rangle$,} the sign of which provides the classification result, i.e., $\tilde{y}=+1$ for normal data and $\tilde{y}=-1$ for fraudulent data.
As test data, we take the top four normal data IDs and fraudulent data IDs, in ascending order, excluding the training data. 
Classification results are shown in {Table}~\ref{tb:fraud_detection}.

\begin{table}[htb]
\caption{The results of fraud detection.}
\centering
    \begin{threeparttable}[h]
    \begin{tabular}{|c|c|c|c|c|} 
    \hline
    Test data ID{\tnote{a}} \quad & Class & $\quad\langle \sigma_z^{(\rm{A})}\rangle\quad$ & $\quad\tilde{y}\quad$ & $\quad$Result$\quad$\\ \hline \hline
    \#5 &  & 0.0699 & $+$ & Correct\\
    \#6 & Normal & 0.1978 & $+$ & Correct\\ 
    \#7 & ($+$) & 0.0183 & $+$ & Correct\\
    \#8 &  & 0.1579 & $+$ & Correct\\ \hline
    \#6330 &  & -0.3747 & $-$ & Correct\\
    \#6332 & Fraud & -0.4245 & $-$ & Correct\\ 
    \#6335 & ($-$) & -0.4289 & $-$ & Correct\\
    \#6337 &  & -0.4163 & $-$ & 
    Correct\\ \hline
    \end{tabular}
    \label{tb:fraud_detection}
    \begin{tablenotes}
    \item[a] As test data, we take the top four normal data IDs and fraudulent data IDs, in ascending order, excluding the training data.
    \end{tablenotes}
    \end{threeparttable}
\end{table}

Note that the instances can be arbitrarily chosen, rather than the top 4 normal and fraudulent data IDs. 
Hence, we have conducted an additional simulation with randomly selected 48 instances chosen from the same Kaggle dataset described above. 
As a result, we have confirmed that 42 out of the 48 test data are correctly classified. 
Although the remaining 6 test data are incorrectly classified, this misclassification may be caused by the intrinsic characteristics of the dataset and/or an insufficient amount of training data rather than the encoding errors.


\section{Conclusion and discussion}
\label{SECTION-conclusion}

In this paper, we proposed the approximate complex amplitude encoding algorithm (ACAE) which allows for the efficient encoding of given complex-valued classical data into quantum states using shallow parametrized quantum circuits. 
The key idea of this algorithm is to use the fidelity as a cost function, which can reflect the difference in complex amplitude between the model state and the target state, unlike the MMD-based cost function in the original AAE.
Also note that the classical shadow with random Clifford unitary is used for efficient fidelity estimation. 
In addition, we applied ACAE to realize the CHC with fewer gates than the original CHC 
which requires an exponential number of gates to prepare the exact quantum state. 
Using this algorithm we demonstrated the Iris data classification and the credit card 
fraud detection that is considered as a key challenge in financial institutions.

The main concern in the use of ACAE is its scalability. 
To discuss this problem, we conducted a numerical simulation for the same credit card fraud detection problem as before \cite{credit}, to see the relationship between the number of data (accordingly the number of qubits) and the circuit depth required to achieve a particular value of fidelity. 
Here we examine the case where the number of data varies from 64 (5 qubits) to 1024 (9 qubits). 
Figure~\ref{FIGURE-depth} shows that the required depth of the variational quantum circuit increases superlinearly with respect to the number of qubits, implying that ACAE may be not applicable to large-size problems. 
However, we found that, for the case of 5 qubits, even when the fidelity is decreased from 0.99+ to approximately 0.7 by reducing the number of training steps, the classification accuracy is still above 80$\%$. 
That is, ACAE may work for some practical problems such that 80$\%$ classification accuracy is enough.

\fighalf{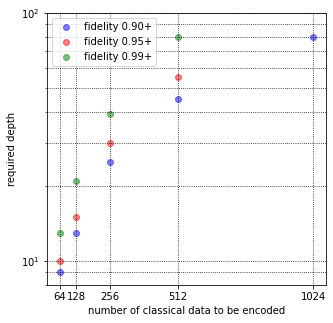}
{The required circuit depth to prepare states that reach the fidelity bigger than 0.90, 0.95, or 0.99. 
For the case of 1024 classical data, we were able to calculate the depth only for the case of fidelity bigger than 0.90, due to the limited computation resources.}{FIGURE-depth}{width=200pt}

For practical applications of ACAE, we need to deal with enormously large datasets. 
For example, in the demonstration for credit card fraud detection, approximately 280,000 training data are provided and each data has 28 different features. This means
$n+m+1=\lceil {\log_2(28)}\rceil + \lceil {\log_2(280,000/2)}\rceil + 1 = 23$ qubits are  required to deal with all the data.
As the number of qubits is increased, the degree of freedom of the quantum state exponentially grows; in such a case, there appear several practical problems to be resolved. 
For example, ACAE employs the fidelity as a cost function, which is, however, 
known as a global cost that leads to the so-called gradient vanishing problem or 
the barren plateau problem \cite{mcclean2018barren}; i.e., the gradient vector of the cost decays 
exponentially fast with respect to the number of qubits, and thus the learning 
process becomes completely stuck for large-size systems. 
To mitigate this problem, recently the localized fidelity measurement has been 
proposed in {References}~\cite{khatri2019quantum,sharma2020noise,tezuka2022generative}. 
Moreover, applications of several existing methods such as circuit initialization \cite{grant2019initialization,zhang2022gaussian}, special structured ansatz \cite{cerezo2021cost,liu2022mitigating}, and parameter embedding \cite{volkoff2021large} are worth investigating to address the gradient vanishing problem. 
Another problem from a different perspective is that the random Clifford measurement 
for producing the classical shadow can be challenging to implement in practice, 
because $\mathcal{O}\left(n^2/{\log_2(n)}\right)$ entangling gates are needed to sample 
from $n$-qubit Clifford unitaries. 
{References}~\cite{maslov2022depth,MW2022} have presented that the Clifford circuit depth over unrestricted architectures is upper bounded by $2n+\mathcal{O}\left({\log_2^2(n)}\right)$ for all practical purposes, which may improve the implementation of the fidelity estimation process.
Overall, algorithm improvements to deal with these problems are all important and 
remain as future works.


\section*{Acknowledgments}
This work was supported by Grant-in-Aid for JSPS Research Fellow No. 22J01501 and MEXT Quantum Leap Flagship Program Grants No. JPMXS0118067285 and No. JPMXS0120319794.
We acknowledge the use of IBM Quantum services for this work. We specially thank Kohei Oshio, Ryo Nagai, Ruho Kondo, Yuki Sato, and Dmitri Maslov for their comments on the early version of this draft. We are also grateful to Shumpei Uno for helpful support and discussions on the early stage of this study.

The views expressed are those of the authors and do not reflect the official policy or position of IBM or the IBM Quantum team.


\bibliographystyle{abb_srt.bst}
\bibliography{main_revised.bib}


\end{document}